\documentclass[aps,prb,superscriptaddress,amsfonts,amsmath,amssymb,floatfix,reprint,longbibliography]{revtex4-2}

\usepackage{url}
\usepackage{bm}
\usepackage{graphicx}
\usepackage{amsmath}
\usepackage{subfigure}
\usepackage{amstext}
\usepackage{amssymb}
\usepackage{amsfonts}
\usepackage{amsbsy}
\usepackage{verbatim}
\usepackage{physics}
\usepackage{braket}
\usepackage{color}
\usepackage[colorlinks=true, urlcolor=blue, linkcolor=blue, citecolor=blue, pdftex]{hyperref}
\usepackage{multirow}
\usepackage{floatrow}
\usepackage{float}
\usepackage{tikz}
\usepackage{gensymb}
\usepackage{textcomp}
\usepackage{enumitem}
\usepackage[version=3]{mhchem}
\usepackage{afterpage}
\usepackage{pbox}
\usepackage{makecell}
\usepackage{dsfont}
\usepackage{physics}
\usepackage{siunitx}
\usepackage{fancyhdr}
\usepackage{stackengine,wasysym,scalerel}
\usepackage{latexsym}
\usepackage{cancel}
\usepackage{array, multirow, bigdelim, makecell, booktabs}
\usepackage[misc]{ifsym}
\usepackage[normalem]{ulem}
\usepackage{times}
\usepackage{comment}
\usepackage{graphicx}

\newenvironment{localgraphicspath}[1]{
  \graphicspath{#1}
}{}

\definecolor{darkblue}{HTML}{004D6B}
\definecolor{darkred}{HTML}{8c1515}
\definecolor{darkgreen}{HTML}{006400}
\usepackage{hyperref}
\hypersetup{
	colorlinks=true,
	urlcolor=darkred,
	citecolor=darkblue,
	linkcolor=darkred,
	breaklinks
}

\begin{document}

\title{Quantum phase diagram and non-abelian Moore-Read state in double twisted bilayer graphene}

\author{Sen Niu}\email{sen.niu@csun.edu}
\affiliation{Department of Physics and Astronomy, California State University Northridge, California 91330, USA}

\author{Yang Peng}\email{yang.peng@csun.edu}
\affiliation{Department of Physics and Astronomy, California State University Northridge, California 91330, USA}
\affiliation{Institute of Quantum Information and Matter and Department of Physics, California Institute of Technology, Pasadena, CA 91125, USA}

\author{D. N. Sheng}\email{donna.sheng@csun.edu}
\affiliation{Department of Physics and Astronomy, California State University Northridge, California 91330, USA}

\date{\today}

\begin{abstract}

Experimental realizations of Abelian fractional Chern insulators (FCIs) have demonstrated the potentials of moir\'e systems in synthesizing exotic quantum phases. Remarkably, 
twisted multilayer graphene system may also host non-Abelian  states competing with  charge density wave  under Coulomb interaction.  Here, through larger scale exact diagonalization simulations, we map out the quantum phase diagram for $\nu=1/2$ system with electrons occupying the lowest moir\`e band of the double twisted bilayer graphene. 
By increasing the system size, we find the ground state has  six-fold near degeneracy and with a  finite spectral gap separating
the ground states from excited states across a broad range of parameters. Further computation of many-body Chern number establish the topological order of the state, and we rule out possibility of charge density wave orders based on featureless density structure factor. Furthermore, we inspect the particle-cut entanglement spectrum to identify the topological state as a non-Abelian Moore-Read state. Combining all the above evidences we conclude that Moore-Read ground state dominates the quantum phase diagram for the double twisted  bilayer  graphene system for a broad range of  coupling strength with realistic Coulomb interaction.

\end{abstract}


\maketitle

\emph{Introduction.---} Band topology and electron correlations are crucial for understanding exotic quantum phases of matter. A pronounced example resulting from interplay between both of them is the fractional quantum hall (FQH) effect, which is observed in two-dimensional electron gas subjected to strong magnetic fields \cite{tsui1982two}. FQH states are topological ordered phases and host elementary quasi-particles with fractionalized charge \cite{laughlin1983anomalous} and fractionalized statistics, dubbed anyons \cite{wilczek1990fractional}. Among the family of anyons in FQH effects, the most fascinating subclass is the non-Abelian anyons which are proposed to have important applications in robust quantum computations against local disturbances \cite{nayak2008non}. As lattice generalizations of FQH states, fractional Chern insulators \cite{neupert2011fractional,sheng2011fractional,regnault2011fractional,tang2011high,sun2011nearly} (FCIs) emerge from correlated topological band structures without the need of strong field and provide a new avenue towards manipulating anyons. 

Over the past few years, moir\'e materials become an exciting platform for exploring zero field FCIs. Misaligning two layers of graphenes at magic twist angles, flat topological Chern bands can be engineered in the moir\'e superlattice \cite{tarnopolsky2019origin} as confirmed by observations of intrinsic quantum anomalous Hall effects with spontaneous spin and valley polarization \cite{serlin2020intrinsic,nuckolls2020strongly} in twisted bilayer graphenes (TBGs). Under moderate magnetic field $B= 5$\text{T} \cite{xie2021fractional}, FCIs can be realized in TBGs as indicated by local compressibility measurement, where the magnetic field just plays the role of improving 
spin and valley polarization  
and destabilizing competing charge density wave (CDW) orders.
More recently, a series of experiments demonstrated the realization of zero field abelian FCIs in both twisted transition metal dichalcogenides (TMDs) \cite{cai2023signatures,zeng2023thermodynamic,park2023observation,xu2023observation} and (non-twisted) rhombohedral pentalayer graphene/hBN mori\'e superlattices \cite{lu2024fractional}.  A possible non-Ablian zero-field FCI was also reported experimentally in rhombohedral hexalayer graphene/hBN systems \cite{xie2024even}, where an even-denominator Hall plateau was observed. Nevertheless, zero-field FCIs have not yet been realized in twisted graphene multilayers despite the similarities to the other two platforms. 

Recent theoretical advances suggest the potential for non-Abelian FCIs in twisted TMDs~\cite{chen2024robust, reddy2024non, xu2024multiple, ahn2024first}. Significant progress has also been made toward realizing these states in graphene-based systems. In particular, Ref.~\cite{fujimoto2024higher} demonstrated that introducing interlayer couplings between two copies of magic-angle TBGs can produce first excited Landau level like Chern bands with a similar quantum geometry.   
A subsequent numerical study \cite{liu2024non} reported signatures of Moore-Read states \cite{moore1991nonabelions} in these models under Coulomb interaction. However, 
it remains an open question  whether Moore-Read states can be stabilized as gapped ground states in the thermodynamic limit, due to the presence of very low-energy  excitations \cite{girvin1986magneto} across different system sizes. 
Consequently, the quantum phase diagram  and the potential for a non-Abelian phase in this double-twisted bilayer system require further systematic studies.

In this letter, we perform large-scale exact diagonalization (ED) calculations to accurately determine the quantum phases of the double magic-angle TBG system. Specifically, we consider the continuum model of coupled TBGs \cite{fujimoto2024higher,liu2024non} within the parameter regime of first Landau level like bands and diagonalize the half-filled ($\nu = 1/2$) many-body Hamiltonian. By increasing the system size, we demonstrate that the six-fold quasi-degenerate states evolve into gapped ground states, with the spectral gap remaining finite over a broad range of coupling strengths and dielectric constants ($\epsilon$). Calculations of the many-body Chern number and structure factor further reveal that the Chern numbers align with the half-quantized Hall conductance of Moore-Read state, ruling out the possibility of CDW orders by the absence of sharp peaks in density structure factor. Finally, an analysis of the particle-cut entanglement spectrum confirms that the wavefunctions follows  the unique exclusion rule of Moore-Read state. Collectively, these findings provide strong evidence that, across a wide range of coupling parameters and under realistic Coulomb interaction, robust Moore-Read ground states exist in the coupled TBG model with no CDW order.

\begin{figure}[t]
\begin{localgraphicspath}{{maintext_fig/}}
\includegraphics[width=0.9\columnwidth]{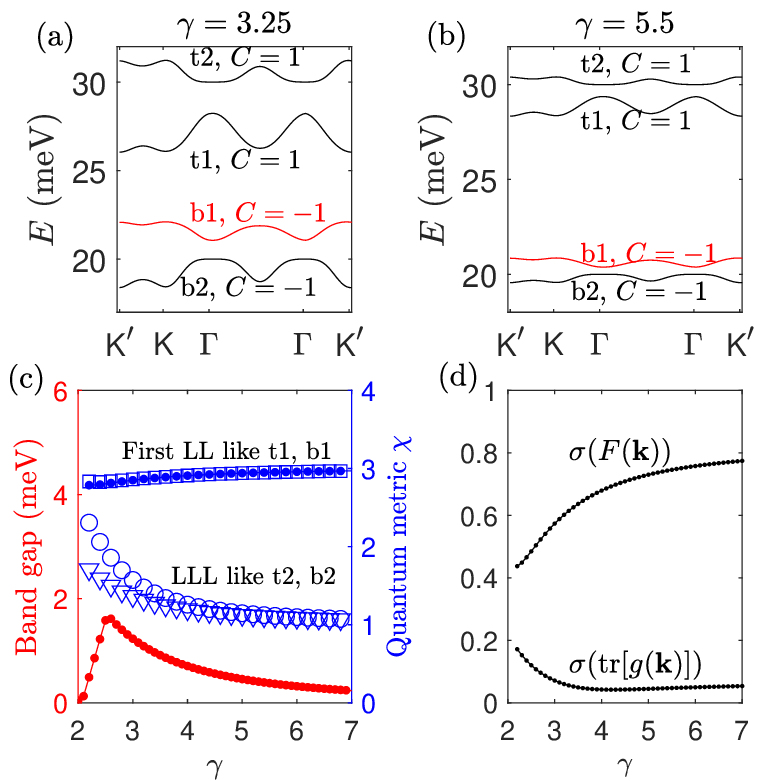}
\caption{\label{fig:band} 
Single particle properties of the continuum model. (a)-(b) band structures and Chern numbers of four middle bands (two top bands t1, t2 and two bottom bands b1, b2) for $\gamma=3.25$ and $5.5$ respectively. (c) quantum metric $\chi$ (blue curves) of four bands. The four bands are fully gapped and we plot band gap for the first bottom band b1 (red curve) as an example. (d) standard deviations of berry curvature and geometric tensor trace for the first b1 band. }
\end{localgraphicspath}
\end{figure}

\emph{Continuum model.---} We consider the continuum model of double magic-angle TBGs \cite{fujimoto2024higher,liu2024non,liu2024theory}, described by the following Hamiltonian
in the chiral basis ordered as $\ket{1A}, \ket{2A}, \ket{1B}, \ket{2B}, \ket{3A}, \ket{4A}, \ket{3B}, \ket{4B}$ (the number denotes the layer index and $A/B$ refers to the two sublattices of graphene)
\begin{align}
H_{0}(\bold r)&=\left(\begin{array}{cccc}
u_{1} & \mathcal{D}_{1}^{*}(-\bold r) & 0 & 0\\
\mathcal{D}_{1}(\bold r) & u_{2} & \Gamma I & 0\\
0 & \Gamma I & u_{3} & \mathcal{D}_{2}^{*}(-\bold r)\\
0 & 0 & \mathcal{D}_{2}(\bold r) & u_{4}
\end{array}\right),
\label{eq:Hamiltonian} \\
\mathrm{with\ }& 
\mathcal{D}_{i}(\bold r)=\left(\begin{array}{cccc}
-2\hbar v_0\bar{\partial} +m_{i,1} & w U(\bold r) \\
 w U(-\bold r) & -2\hbar v_{0}\bar{\partial} +m_{i,2} \\
\end{array}\right) \nonumber,
\end{align}
where $\bar{\partial}=\frac{1}{2}(\partial_x+i\partial_y)$, $I$ is the identity submatrix and $U(\bold r)=e^{-i\bold{q}_1 \bold{r}}+e^{i\phi}e^{-i\bold{q}_2 \bold{r}}+e^{-i\phi}e^{-i\bold{q}_3 \bold{r}}$ is the interlayer tunneling potential in TBG \cite{bistritzer2011moire}. When the coupling strength $\Gamma=0$ and potentials $m_{ij}=0,u_i=0$, Eq.~\eqref{eq:Hamiltonian} reduces to two copies of decoupled chirally symmetric TBGs \cite{tarnopolsky2019origin}, i.e., layer pairs $1,2$ and $3,4$. We take parameters of the first magic angle $\theta\approx 1.09^{\circ}$ with $w=110\text{meV}$, $\alpha=w/\hbar v_{0} k_{\theta}\approx 0.586$, $k_\theta=2k_D \sin(\theta/2)$ and $k_D$ being reciprocal lattice spacing of graphene. In such a decoupled limit, the four flat Chern bands in the middle of energy spectrum are exactly degenerate. We take the sublattice potential parameters $(u_1,u_2,u_3,u_4)=(20,-20,50,30)\text{meV}$ to split the four Chern bands and $(m_{11},m_{12},m_{21},m_{22})=(20,0,0,20)\text{meV}$ which slightly modify band widths. The $\Gamma$ term which couples different sublattices between next nearest neighbour layers is responsible for appearance of higher Landau level physics. For simplicity we use a dimensionless parameter $\gamma=\Gamma\alpha/w$ to label strength of $\Gamma$. We note such model can also be viewed as a three-dimensional alternatively AB-BA twisted Bernal graphene when projected to $k_z=0$ component \cite{fujimoto2024higher}.

The features of the moir\'e topological bands can be analyzed by comparing with Landau levels. In ideal Landau levels, the Chern number is always $C=1$ while the quantum metric $\chi_n=\frac{1}{2\pi}\int d\mathbf{k}\text{tr}(g_n(\mathbf{k}))=2n-1$ depends on the Landau level index $n$ \cite{ozawa2021relations}, where $g_n(\mathbf{k})$ is the $2\times 2$ quantum geometric tensor \cite{provost1980riemannian}. Turning back to our continuum model, without interlayer coupling $\gamma$, the four middle bands inherit Chern numbers $C=(+1,+1,-1,-1)$ of the magic TBG with quantum metric being $\chi= 1$. For nonzero $\gamma$, typical band structures are shown in (Fig.~\ref{fig:band} (a)-(b)), where the the Chern numbers remain unchanged. However, from the quantum metric of the four band in the fully gapped regime (Fig.~\ref{fig:band} (c)), one can see for $\gamma\ge 3$ the first top/bottom band become first Landau level like since $\chi\approx 3$.  
From the standard deviations in Fig.~\ref{fig:band} (d) for the b1 band, one can see that the geometric tensor trace 
$\text{tr}(g(\mathbf{k}))$ is relatively uniform while the Berry curvature variance only has a small increase with increasing $\gamma$. 
 We will consider FCI states realized in the partially filled first bottom (b1) band and present  similar results for the first top (t1) band  in Supplemental Materials (SM) \cite{SM}.

We now include Coulomb interaction which is necessary for emergence of fractional quantum Hall effects
\begin{align}
H_{int}=\frac{1}{2A}\sum_{n_{1,2,3,4}} & \sum_{\bold{k}_1\bold{k}_2\bold{q}}  V_{n_1,n_2,n_3,n_4}(\bold{k}_1, \bold{k}_2,\bold{q}) \times  \notag \\
&\psi_{n_1,\bold{k}_1}^{\dagger}\psi_{n_2,\bold{k}_2}^{\dagger}\psi_{n_3,[\bold{k}_2-\bold{q}]}\psi_{n_4,[\bold{k}_1+\bold{q}]},
\label{eq:Coulomb}
\end{align}
where $A=\frac{8\pi^2 N_s}{3\sqrt{3} k_{\theta}^2}$ is the cluster area and $N_s$ the number of sites.
The summation of $\bold{k}_1,\bold{k}_2$ is over the moir\'e first Brillouin zone (FBZ), $n_{i}$ labels the band index, $[k-q]$ denotes the reduced momentum in FBZ which is shifted by multiples of moir\'e reciprocal lattice vectors, and 
$V_{n_1,n_2,n_3,n_4}(\bold{k}_1, \bold{k}_2,\bold{q})=V(\bold{q})\langle n_1,\bold{k}_1|e^{-i\bold{q} \cdot \bold{r}}|n_4,[\bold{k_1}+\bold{q}]\rangle \langle n_2,\bold{k}_2|e^{i\bold{q} \cdot \bold{r}}|n_3,[\bold{k_2}-\bold{q}]\rangle$ is evaluated from overlap matrix between Bloch states. $V(\bold q)=\frac{2\pi e^{2}k_0}{|\bold q| \epsilon}(1-\delta_{\bold{q},0})$ is fourier transformed interaction, where $k_0$ is Coulomb constant and the dielectric constant $\epsilon$ controls interaction strength. 
For numerical simulation of the many-particle Fock space, we adopt the single-band approximation and project the Coulomb interaction onto the b1 band that we are interested in without considering Hartree-Fock energy contributed from fully-filled lower bands. Such projection allows us to treat the many-body Hamiltonian $H=H_{0}+H_{int}$ defined on finite-sized clusters up to $N_s=32$ sites using exact diagonalization method. Note that different from Ref. \cite{liu2024non} which ignores band dispersion,
we include $H_0$ term and address the problem under realistic Coulomb interaction strength comparable with band width. On finite clusters, the allowed discrete momentum grid takes the form $\bold{k}=k_1 \mathbf{T}_1+k_2 \mathbf{T}_2$. Here $\mathbf{T}_{1(2)}$ are unit momentum vectors, and $k_{1(2)}=0,1,2,...,N_{1(2)}-1$ labels the coordinate on the $N_s=N_{1}N_{2}$ momentum space lattice. Symmetries and qualities of all the clusters we use are summarized in SM. The total momentum of occupied electrons $\mathbf{k}=\sum_{i=1}^{N_p}\mathbf{k}_i$ is used as a good quantum number to block diagonalize the many-body Hamiltonian in Fock space with filling fraction $\nu=N_p/N_s=1/2$.

\begin{figure}[t]
\begin{localgraphicspath}{{maintext_fig/}}
\includegraphics[width=0.9\columnwidth]{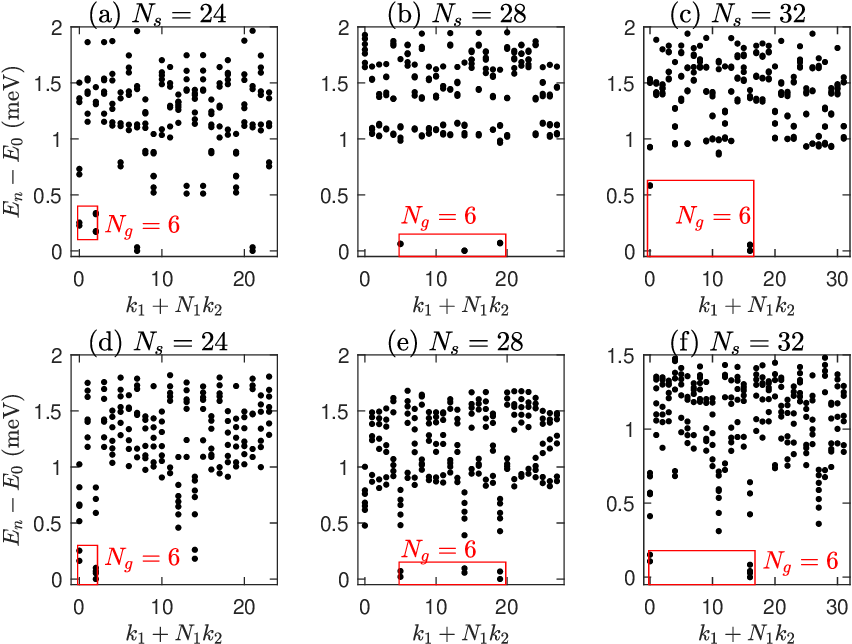}
\caption{\label{fig:ED_spectrum} 
Momentum resolved many-body spectrum at $\epsilon=6$ for $\gamma=3.25$ in (a)-(c) and $\gamma=5.5$ in (d)-(f). }
\end{localgraphicspath}
\end{figure}

\begin{table}[t]
\begin{tabular}{|c|c|c|c|c|c|}
\hline 
$N_{s}$ &  & $(0,0)$ & $(0,\pi)$ & $(\pi,0)$ & $(\pi,\pi)$\tabularnewline
\hline 
\hline 
24 & \multirow{3}{*}{$k_{1}+N_{1}k_{2}$ } & 0 & 14 & 2 & 12\tabularnewline
\cline{1-1}\cline{3-6}
28 &  & 0 & 19 & 5 & 14\tabularnewline
\cline{1-1}\cline{3-6}
32 &  & 0 & 16 & 27 & 11\tabularnewline
\hline 
24 & \multirow{3}{*}{Degeneracy} & 2 & 0 & 4 & 0\tabularnewline
\cline{1-1}\cline{3-6}
28 &  & 0 & 2 & 2 & 2\tabularnewline
\cline{1-1}\cline{3-6}
32 &  & 2 & 4 & 0 & 0\tabularnewline
\hline 
\end{tabular}
\caption{\label{table:momentum counting} Momentum counting of the $6$ fold Moore-Read states on different clusters derived from generalized pauli principle . The first three and second three rows show the 1D index $k_1+N_1 k_2$ and distributions of $6$ fold states, respectively.}
\end{table}

\emph{Many-body energy spectrum.---}We investigate the many-body phases in the regime  $3\le\gamma\le 7$, where the band projection approximation can be justified by a reasonable large  single-particle band gap in Fig.~\ref{fig:band} (c) for the b1 band. 
To identify the many-body phase, we analyze the energy spectrum, including ground state degeneracy and the existence of energy gap. The topological degeneracy of Moore-Read state can be derived from the root configurations after mapping momentum states to  1D  orbitals
\cite{bergholtz2006pfaffian,seidel2006abelian,read2006wavefunctions,ardonne2008degeneracy,haldane1991fractional,regnault2011fractional,bernevig2012emergent}, where the three-body exclusion rule imposes the constraint that on any four adjacent orbitals no more than two particles can be occupied. For $N_p=2n$ only periodic configurations like $10101010...$ and $11001100...$ are allowed while for $N_p=2n+1$ only the former is allowed. Thus the ground state degeneracy $N_g=6$ for even electron number and $N_g=2$ for odd electron number.
We focus on clusters of even particle number with $N_p=2n$ ($N_s=4n$) since  moore-Read states are paired states \cite{read2000paired} and present 
$N_p=2n+1$ cluster results in SM.

To identify Moore-Read states on different cluster geometries, we determine the momentum quantum numbers $\mathbf{k}$ through folding the $N_s$ single particle orbitals in 1D order $k_1+N_1 k_2$\cite{haldane1991fractional,regnault2011fractional,bernevig2012emergent}. We get the momentum countings of the six 1D occupation configurations mentioned above and display the corresponding momentum resolved degeneracies in Table \ref{table:momentum counting}. The energy spectrum for $N_s=24,28,32$ clusters are shown in Fig.~\ref{fig:ED_spectrum} with different $\gamma$ values and fixed dielectric constant $\epsilon=6$.  For all clusters we observe characteristic $N_g=6$-fold quasi-degeneracy as shown in Fig.~\ref{fig:ED_spectrum}. These low-energy multiplets emerge as lowest levels in correct momentum sectors with correct degeneracy as predicted in Table \ref{table:momentum counting}, which is the first signature of Moore-Read states. We notice that for small cluster $N_s=24$ which has asymmetric aspect ratio $1.5$ \cite{SM}, the spectrum is not fully gapped due to existence of lower energy states in other $\mathbf{k}$ sectors. 
However, the situation changes by looking into larger clusters $N_s=28,32$. For the $N_s=28$ cluster with $C_6$ rotation symmetry,  one can see clearly $6$-fold multiplets as ground states separated from high energy excitations. Moreover, on the largest $N_s=32$ cluster (despite there is no $C_6$ symmetry for this cluster), there is still a finite excitation gap indicating the strengthening of the Moore-Read phase with increasing system sizes. The energy spectra are similar for different $\gamma$ values. These evidences suggest that the $6$-fold multiplets are true ground states in the thermodynamic limit.

\begin{figure}[t]
\begin{localgraphicspath}{{maintext_fig/}}
\includegraphics[width=0.9\columnwidth]{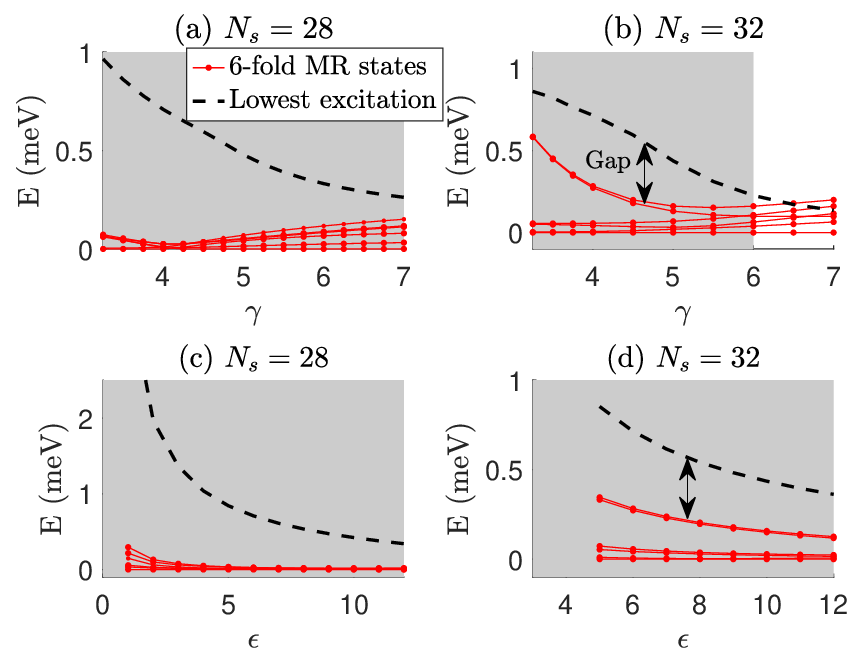}
\caption{\label{fig:manybody_spectrum} 
Many-body spectrum versus $\gamma$ or $\epsilon$. Red and black curves correspond to $6$-fold quasi-degenerate Moore-Read states and lowest excitations.   $\epsilon=6$ in (a)-(b) and $\gamma=4$ in (c)-(d). The parameter regimes of gapped Moore-Read ground states are shaded.}
\end{localgraphicspath}
\end{figure}

To determine the region of the gapped phase, we plot the low lying energies of the six Moore-Read states as well as lowest  excitation energy on large clusters $N_s=28,32$ in Fig.~\ref{fig:manybody_spectrum} (a)-(b). 
Remarkably, unlike the case in Fig.~\ref{fig:ED_spectrum} (a) and (d) with $N_s=24$,  the $6$ fold states win the competition with the low-energy excited states in other momentum sectors and become fully gapped ground states through the range of $3\le \gamma \le 6$. Although Ref.~\cite{liu2024non} proposed a quantum phase transition between Moore-Read state and CDW state around $\gamma_c\approx 4.9$ by showing a local minimum of spectral gap on the $N_s=20$ cluster  \cite{SM}, such possibility is ruled out by our results on larger $N_s=28,32$ clusters. 
We further examined the static structure factors $S(\mathbf{q})$ in SM, and find that the $S(\mathbf{q})$ is similar for $\gamma=3.25$ and $\gamma=5.5$ without sharp peaks in moir\'e Brillouin zone. On the other hand, the interaction strength also matters since band width is nonzero. From the plot of spectrum versus Coulomb dielectric constant $\epsilon$ in Fig.~\ref{fig:manybody_spectrum} (c)-(d) one can also see sizable spectral gap for a wide range of $\epsilon=1-12$, further demonstrating robustness of such many-body phase under realistic Coulomb interaction.

\begin{figure}[t]
\begin{localgraphicspath}{{maintext_fig/}}
\includegraphics[width=0.9\columnwidth]{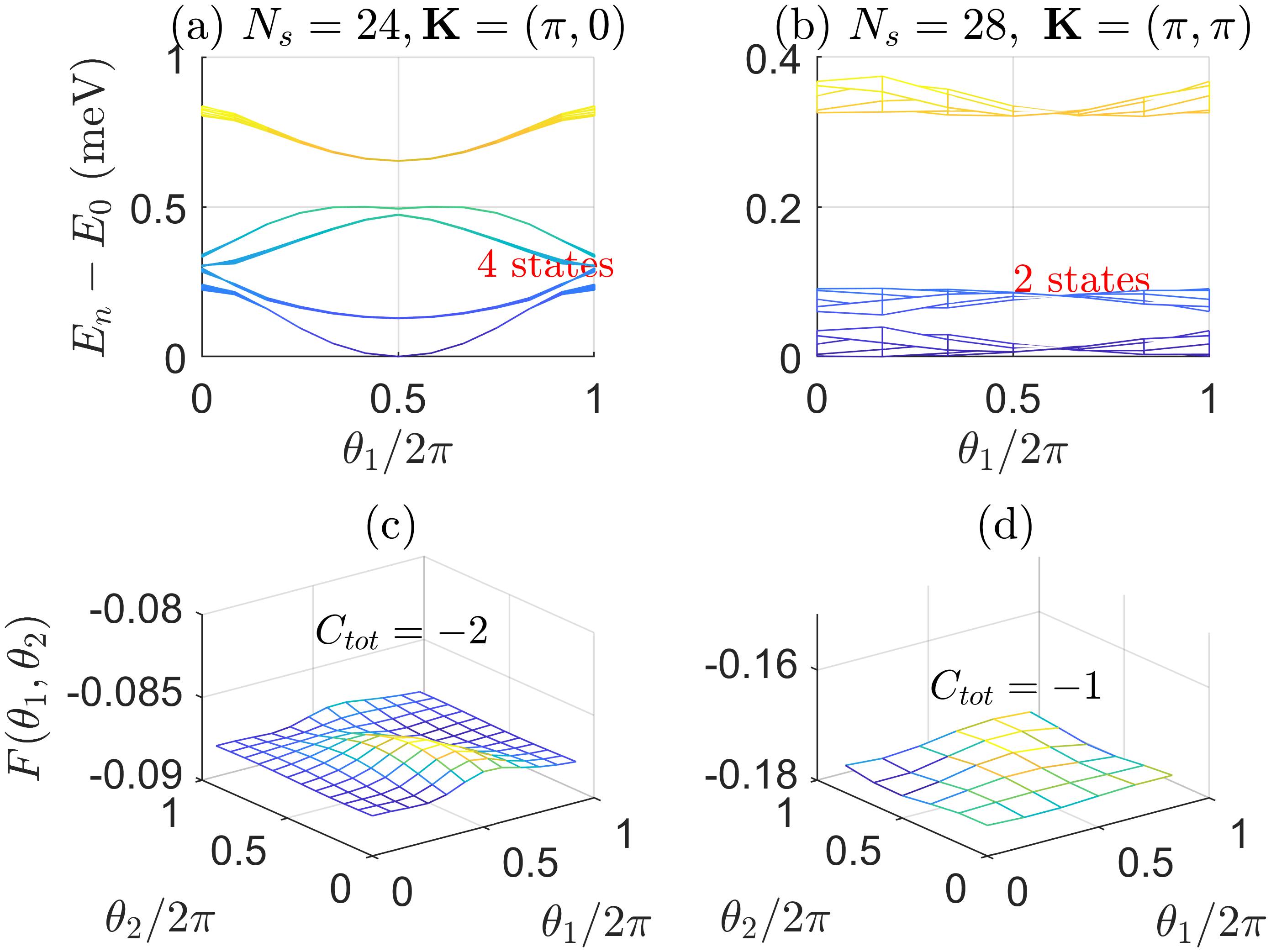}
\caption{\label{fig:flux_spectrum} 
Many-body spectrum flow (main view) under insertion of flux $(\theta_1,\theta_2)$ and distribution of many-body berry curvature $F(\theta_1,\theta_2)$ for $\gamma=5.5,\epsilon=6$. Only representative momentum sectors are displayed. In (a)-(b), the evolution of spectrum along direction $\theta_2$ is hidden but can be inferred from the width of the curves. In (c)-(d), $C_{mean}=-0.5$ for each state below the gap.}
\end{localgraphicspath}
\end{figure}

\emph{Many-body Chern number.---}
We next compute many-body Chern number \cite{niu1985quantized,sheng2003,fukui2005Chern} as an important evidence to support topological nature of ground states. The Chern number for a single many-body ground state $|\psi\rangle$ is defined as an integral over twisted boundary conditions
\begin{align}
C=\frac{1}{2\pi}\int_{0}^{2\pi}d\theta_1 \int_{0}^{2\pi}d\theta_2 F(\theta_1,\theta_2),
\label{eq:Chern}
\end{align}
where the many-body berry curvature takes form $F(\theta_1,\theta_2)=i(\langle \partial_{\theta_1}\psi|\partial_{\theta_2}\psi \rangle-\langle \partial_{\theta_2}\psi|\partial_{\theta_1}\psi \rangle)$ and twisted boundary conditions $\theta_1,\theta_2$ play the role of shifting the discrete momentum grid.
The Many-body Chern number is connected to the Hall transport, and predicts Hall conductance $\sigma_{H}=\frac{e^2}{h}C_{mean}$ with $C_{mean}=C_{tot}/N_g$ and $C_{tot}$ is the sum of Chern numbers over degenerate ground states.

We show examples of spectral flow and distribution of berry curvatures for $\gamma=5.5$ in Fig.~\ref{fig:flux_spectrum}.
With varying $\theta_1, \theta_2$, the energy spectrum gap between near degenerating  ground states and excited states
remains finite (Fig.~\ref{fig:flux_spectrum} (a)-(b)) and the many-body berry curvature remains relatively flat (Fig.~\ref{fig:flux_spectrum} (c)-(d)), which suggests robustness of spectral gap under flux insertion. For $N_s=24$ and $28$, we show one representative momentum sector for each cluster, where the $C_{tot}$ are $-2$ and $-1$ with the degeneracies $N_g$ being $4$ and $2$, respectively, yielding an averaged Chern number $C_{mean}=-0.5$ for each state. Here the minus sign comes from the minus Chern number of the b1 band. When further  examining on all clusters $N_s\ge 24$ (as well as all momentum sectors predicted in Table \ref{table:momentum counting}) at different coupling strengths $\gamma=3.25$ and $5.5$,  we always find half-quantized Chern number $C_{mean}=-0.5$ which is consistent with the theoretical expectation 
for a fractional Chern insulating phase at half-filling of the energy band.  Such observation confirms the  gapped phase as an even denominator FCI state. 

\begin{figure}[t]
\begin{localgraphicspath}{{maintext_fig/}}
\includegraphics[width=0.9\columnwidth]{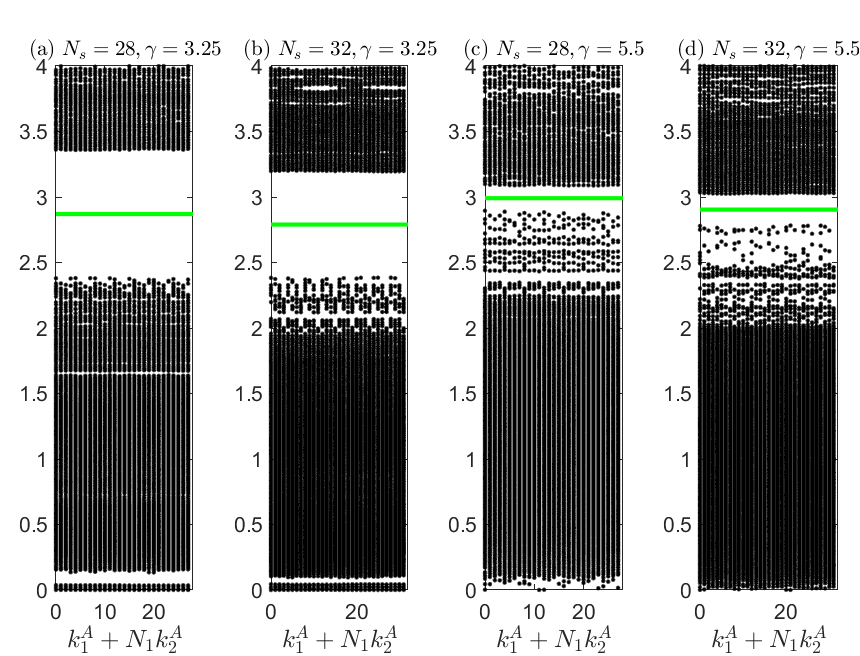}
\caption{\label{fig:PES} 
Momentum resolved PES for bipartition $N_A=4$ with $\epsilon=6$. The green lines denotes the counting for Moore-Read state derived from generalized Pauli principle.}
\end{localgraphicspath}
\end{figure}

\emph{Particle-cut entanglement spectrum.---}
To further determine the topological nature of the many-body ground states, we compute the particle-cut entanglement spectrum (PES) \cite{li2008entanglement,regnault2011fractional,sterdyniak2011extracting}, which are eigenvalues of the reduced density matrix $\rho_A=\text{tr}_B[\sum_{i=1}^{N_g}|\psi_i\rangle\langle|\psi_i|]$ with partitioning the system into two parts with particle numbers $N_A$ and $N_B$,  $N_A+N_B=N_p$ and the sum is over $N_g$ degenerate ground states. It has been shown that the PES can be used to extract quasiparticle statistics derived from generalized Pauli principle \cite{sterdyniak2011extracting}, which requires that no more than two electrons can be occupied for any four consecutive orbitals for Moore-Read states. Thus the characteristic PES counting can be viewed as hallmark of Moore-Read states. The results for $N_A=4$ are displayed in Fig.~\ref{fig:PES}. For $\gamma=3.25$, we observe a spectrum gap which separates larger PES levels from    lower levels (Fig.~\ref{fig:PES} (a)-(b)) for $N_s=28, 32$. This implies that the subsystem configurations that satisfy the generalized pauli principle have much larger weight in the global wavefunction. On the side of $\gamma=5.5$, the PES gap decreases (Fig.~\ref{fig:PES} (c)-(d)) but still remains finite and visible, indicating we have the same quantum phase for both $\gamma$ points, consistent with many-body energy spectra. 
It is interesting to see that among the proposed twisted TMD models \cite{chen2024robust,reddy2024non} as candidates for realizing Moore-Read states, the largest subsystem size $N_A$ satisfying expected Moore-Read PES gap is often around  $N_A=3$, with reduced or closing of spectrum  gap at $N_A=4$. 
Hence the double TBG model we study shows relatively strong signals from PES spectrum for Moore-Read ground states for a broad range of coupling $\gamma$.

\emph{Summary.---} We have performed a comprehensive study on a microscopic continuum model of twisted multilayer graphene and identified Moore-Read ground states in a broad parameter regime under realistic Coulomb interaction. 
Through large scale exact diagonalization of the many-body Hamiltonian on larger clusters, we identify that six fold fully gapped ground states are stabilized and developed, which are distinctly different from  smaller cluster spectra with gapless excitations. The topological properties are investigated through calculating many-body Chern number and particle-cut entanglement spectrum, which all match with the non-Abelian Moore-Read states. 
In particular, for the large interlayer coupling regime with slightly larger  berry curvature fluctuations, we establish
the Moore-Read topological order, and exclude the possible existence of CDW order proposed in an earlier study~\cite{liu2024non}.
Our work demonstrates that graphene based twisted materials are promising candidates for experimental discovery
of exotic non-Abelian FCIs.

\section*{Acknowledgements}
Sen Niu thanks Hui Liu for helpful discussions about the construction of the continuum model and computation of PES. This work is supported by the US National Science Foundation (NSF) Grant No. PHY-2216774.

\bibliography{manuscript.bib}

\clearpage

\appendix
\begin{center}
\title{Supplemental Materials}
\maketitle
\end{center}

\renewcommand\thefigure{\thesection S\arabic{figure}}
\renewcommand\theequation{\thesection S\arabic{equation}}
\setcounter{figure}{0} 
\setcounter{equation}{0}

\section{Finite clusters for exact diagonalization}
We denote $\mathbf{a}_1,\mathbf{a}_2$ as the lattice vectors of the moiré trianglular superlattice. The reciprocal lattice vectors takes the form
\begin{align}
\mathbf{g}_i=\frac{2\pi\epsilon_{ij}\mathbf{a}_j\times\hat{z}}{|\mathbf{a}_1\times\mathbf{a}_2|}.
\end{align}
A finite cluster can be determined from real space translation vectors $\mathbf{L}_1=m_1 \mathbf{a}_1+n_1 \mathbf{a}_2$, $\mathbf{L}_2=m_2 \mathbf{a}_1+n_2 \mathbf{a}_2$, where vectors $\mathbf{L}_1,\mathbf{L}_2$ define the translation invariance of the finite cluster on infinite lattice. The allowed plane wave momenta takes the form
\begin{align}
\mathbf{T}_i=\frac{2\pi\epsilon_{ij}\mathbf{L}_j\times\hat{z}}{|\mathbf{L}_1\times\mathbf{L}_2|},
\end{align}
which satisfies $\mathbf{L}_i \cdot \mathbf{T}_j =2\pi \delta_{ij}$. The discrete momentum points takes the form of integer combination $\mathbf{k}=k_1 \mathbf{T}_1+k_2\mathbf{T}_2$, where $k_i=0,1,...,N_i-1$ and $N_1 N_2=N_s$. In our many-body calculation, we use total momentum of occupied electrons $\mathbf{k}=\sum_{i=1}^{\nu N_s}\mathbf{k}_i$ as a symmetry to block diagonalize many-body Hamiltonian. For each cluster the total number of $\mathbf{k}$ sectors equals to the cluster size $N_s$. 

In our ED calculation, the adopted clusters which are partly imported from Ref.\cite{wilhelm2021interplay} are listed in Table \ref{table:clusters}. We observed that the quality of clusters is important for finite size analysis, as the asymmetric clusters (e..g, $N_s=24$) underestimates the stability of Moore-Read states. Among all clusters, $N_s=28,30, 32$ have good aspect ratios and/or higher point group symmetries.

\begin{table}
\begin{tabular}{|c|c|c|c|c|c|c|}
\hline
  $N_s$ & $\mathbf{L}_1$ & $\mathbf{L}_2$ & $N_1$ & $N_2$ & Aspect ratio & Symmetry\tabularnewline
\hline
  20 & $(4,-4)$ & $(0,5)$ &4&5& 1.25 & $C_2$\tabularnewline
\hline
  24 & $(4,-4)$ & $(0,6)$ &4&6& 1.5 & $C_2$\tabularnewline
\hline
  26 & $(0,13)$ & $(2,4)$ &13&2& 1.1547 & $C_2$\tabularnewline
\hline
28 & $(-2,6)$ & $(4,2)$ &2&14& 1 & $C_6$\tabularnewline
\hline
30   & $(5,-5)$&$(3,3)$ &5&6&1.0392&$D_2$\tabularnewline
\hline
32 & $(2,4)$ & $(6,-4)$ &2&16& 1 & $D_2$\tabularnewline
\hline
\end{tabular}\caption{\label{table:clusters} Information about finite clusters use in ED.}
\end{table}

\section{Static structure factor}
Here we show static structure factors for even number of electrons to rule out spontaneous charge density wave orders. We calculate the (connected) static structure factor
\begin{align}
S(\mathbf{q})=\frac{\langle \rho_{\mathbf{q}} \rho_{-\mathbf{q}}\rangle - \langle \rho_{\mathbf{q}} \rangle \langle \rho_{-\mathbf{q}}\rangle}{N_s},
\label{eq:Sq}
\end{align}
where $\rho_{\mathbf{q}}=\sum_{\mathbf{k}}c_{\mathbf{k}}^\dagger c_{\mathbf{k}+\mathbf{q}}$ is the momentum transfer operator and $\rho_{\mathbf{q}} \approx   \sum_{k\in \text{FBZ}} \langle u(k)|u(k+q)\rangle  \psi_{\mathbf{k}}^\dagger \psi_{\mathbf{k}+\mathbf{q}}$ after band projection.

\begin{figure}
\begin{localgraphicspath}{{appendix_fig/}}
\includegraphics[width=1\columnwidth]{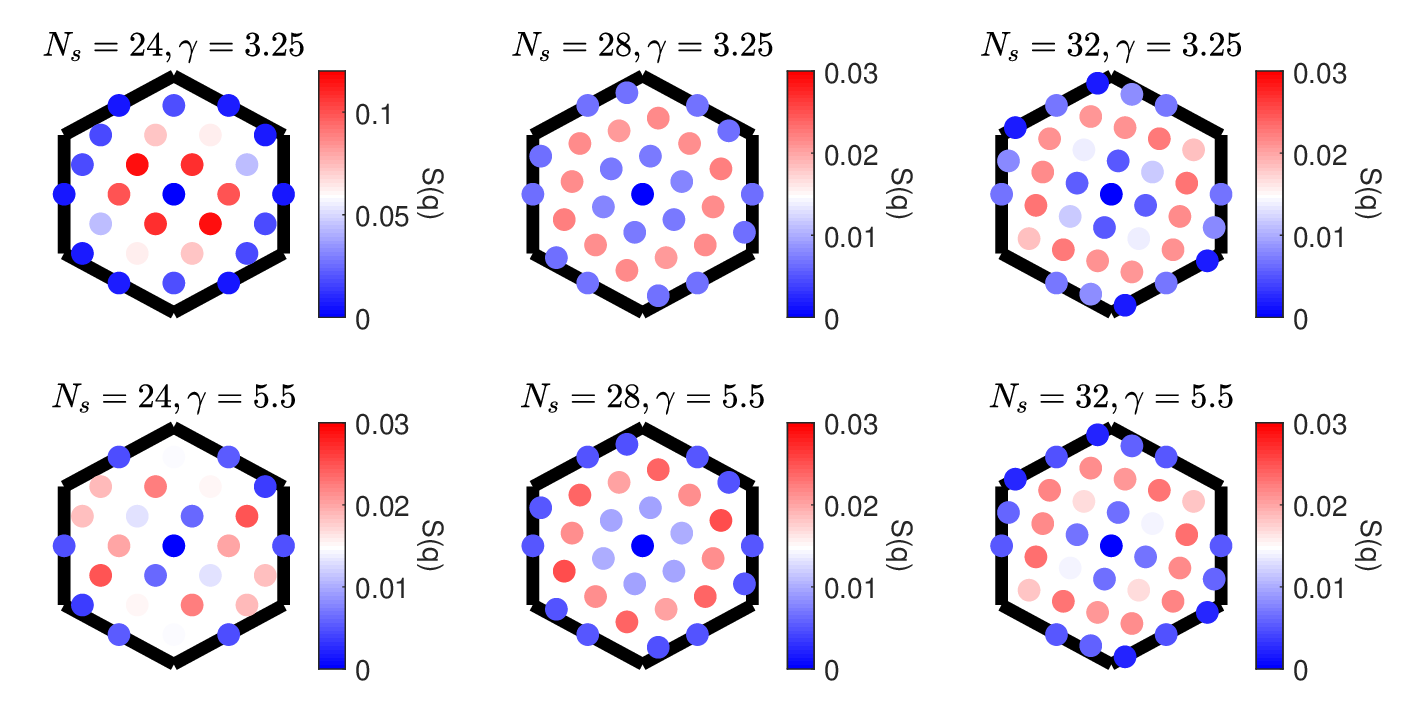}
\caption{Structure factor $S(q)$ computed at $\epsilon=6$. Black dashed line shows moiré Brillouin zone. Absence of sharp singularity indicates no CDW order.   }
\label{Sq}
\end{localgraphicspath}
\end{figure}

The results of structure factor $S(\mathbf{q})$ for even number of electrons are displayed in Fig.\ref{Sq}. The absence of sharp singular points suggests that there is no charge density wave order in these $6$-fold degenerate states. For all the clusters and different parameters, only weak feature of $S(\mathbf{q})$ can be seen: the strength is weak at center and boundary of moiré Brillouin zone, similar to the finding in Ref.\cite{chen2024robust}. Besides, we also checked momentum distribution $n(\mathbf{k})$ in moiré Brillouin zone, which is quite uniform and featureless.


\section{Finite size effect on small clusters}
\begin{figure}
\begin{localgraphicspath}{{appendix_fig/}}
\includegraphics[width=1\columnwidth]{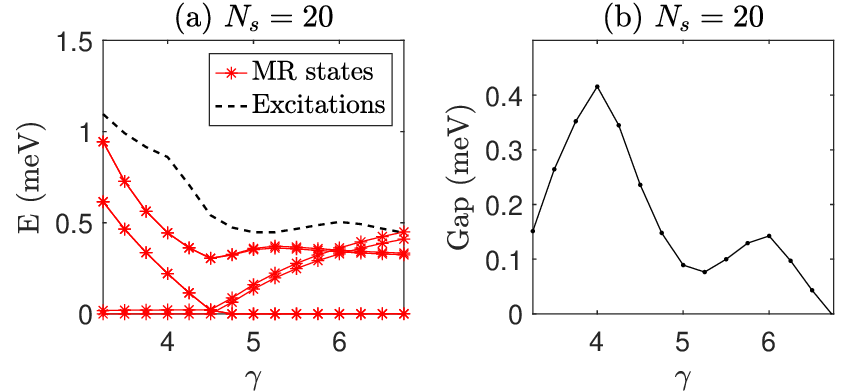}
\caption{Spectral properties on the smaller $N_s=20$ cluster. (a) The $6$-fold Moore-Read states and lowest excitation. (b) The sectral gap versus $\gamma$, which shows a local minimum around $\gamma_c \approx 5.1$.  }
\label{fig:spectrum_20}
\end{localgraphicspath}
\end{figure}

There exists strong finite size effect on phase diagram when cluster size is small. In Fig.\ref{fig:spectrum_20} we show the $6$-fold Moore-Read states and the spectrum gap as functions of $\gamma$. One can see that around $\gamma_c \approx 5.1$, there is a local minimum of the gap, which is also reported by Ref.~\cite{liu2024non} around $\gamma_c \approx 4.9$. Such little quantitative difference is due to neglected band dispersion in Ref.~\cite{liu2024non}. However, as is shown in main text, the results on larger clusters $N_s=28,32$ demonstrate that such gap local minima is an artifact of small cluster size.

\section{Results for odd number of electrons}
In main text, we have analyzed physical properties on clusters with even number of electrons to demonstrate existence of robust Moore-Read ground states. Here we supplement results on clusters with odd number of electrons.
\subsection{Energy spectrum}
For clusters with odd electron numbers, $N_s=4n+2$. The energy spectrum behaves differently from $N_s=4n$. In the thin torus limit (one dimensional periodic chain), one can find only two configurations $101010...$ and $010101...$ are allowed, which indicates a two-fold ground state degeneracy.  The other configurations like $1100110010$ violates the generalized pauli rule since three particles are occupied in the first two and last two orbitals. 

\begin{figure}
\begin{localgraphicspath}{{appendix_fig/}}
\includegraphics[width=0.9\columnwidth]{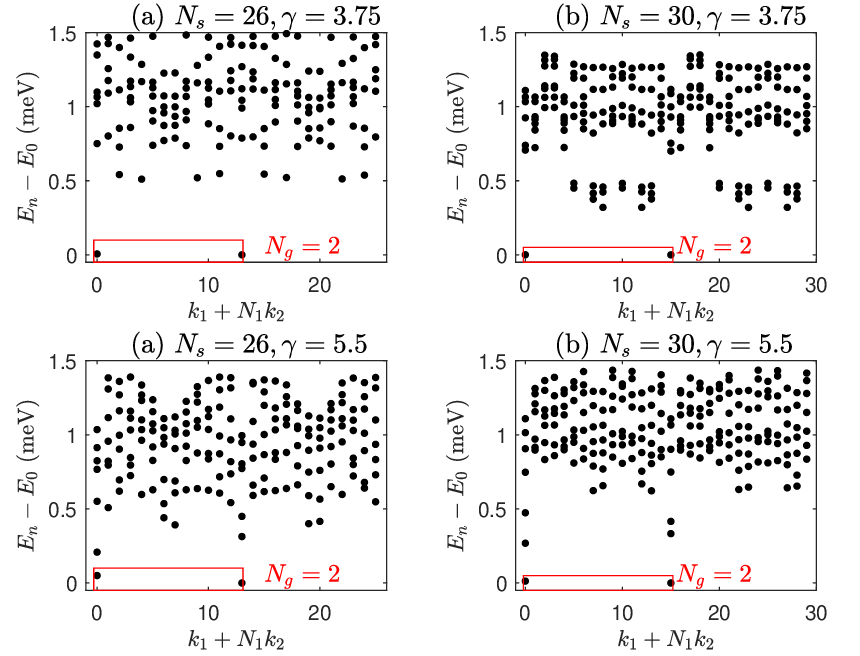}
\caption{Momentum resolved energy spectrum for $\epsilon=6$ on the odd electron number clusters. The momentum quantum number and degeneracies agree with prediction from Moore-Read states.}
\label{fig:spectrum_26_30}
\end{localgraphicspath}
\end{figure}

The energy spectrum for $N_s=26,30$ clusters are shown in Fig.\ref{fig:spectrum_26_30}. There are always two-fold quasi-degeneracy states located at high symmetry points $\mathbf{k}=(0,0)$ and $\mathbf{k}=(0,\pi)$, consistent with theoretical prediction.  Through the whole region $3.25\le\gamma\le7$ the spectrum is fully gapped as can be read from shaded area.

\begin{figure}
\begin{localgraphicspath}{{appendix_fig/}}
\includegraphics[width=0.9\columnwidth]{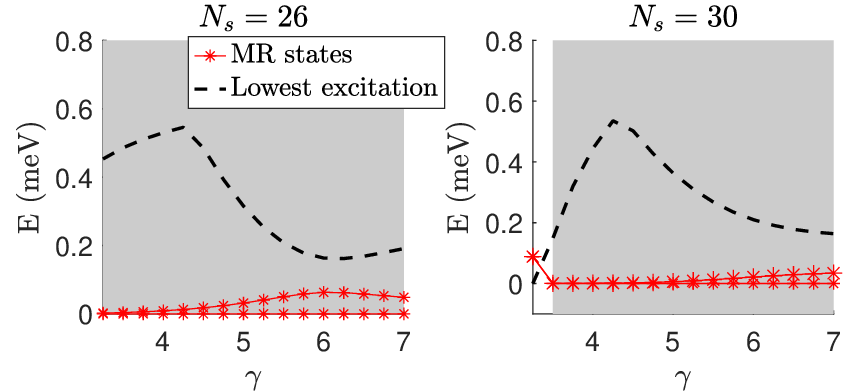}
\caption{Many-body spectrum versus $\gamma$ for $\epsilon=6$ on the odd electron number clusters. Red and black curves correspond to $2$-fold quasi-degenerate Moore-Read states and loest excitations, respectively. The parameter regimes of gapped Moore-Read ground states are shaded. }
\label{fig:manybody_spectrum_26_30}
\end{localgraphicspath}
\end{figure}

\subsection{Many-body Chern number}

The many-body Chern numbser is computed through integration of berry curvature under flux insertion. For nonzero flux, the definition of discrete momentum need to be generalized. The insertion of flux in real space is equivalent to shifting the single-particle momentum grid, which is quantified by $\mathbf{k}(\theta_1,\theta_2) =\mathbf{k}(0,0)+\frac{\theta_1 \mathbf{T}_1+\theta_2 \mathbf{T}_2}{2\pi}$ as a function of inserted flux $\theta_1,\theta_2$. After inserting a single period with $\theta_i$, the global momentum grid is shifted by a single vector $\mathbf{T}_i$ thus the Hamiltonian evolves back to itself, while the total momentum might be changed or unchanged, depending on details of clusters.

\begin{figure}
\begin{localgraphicspath}{{appendix_fig/}}
\includegraphics[width=1\columnwidth]{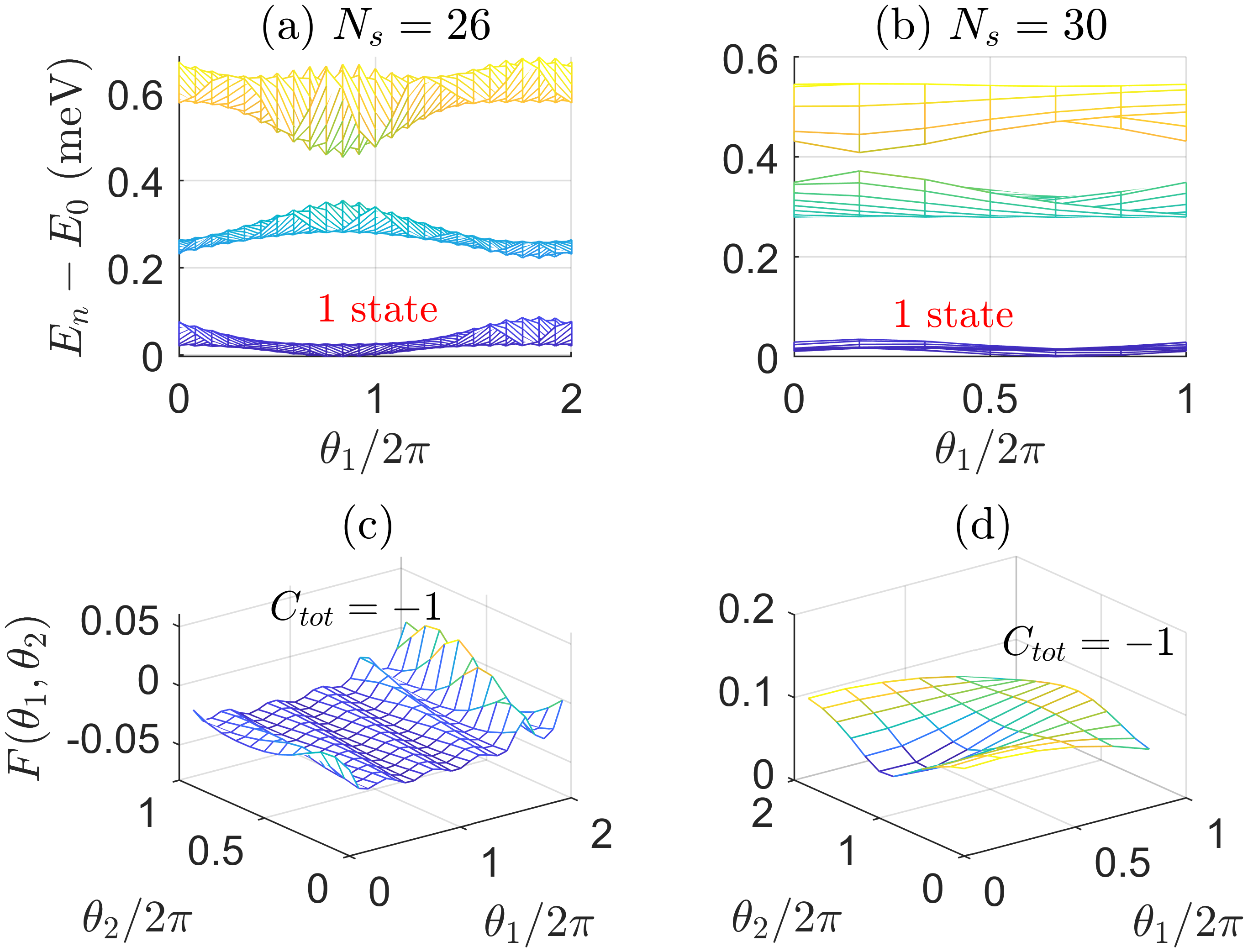}
\caption{Many-body spectrum under flux insersion (a)-(b) and berry curvature (c)-(d) for odd electron numbers with $\gamma=5.5, \epsilon=6$. A single ground state momentum sector evolves back to itself over two flux periods. Integration is over $\theta_1\in [0,4\pi],\theta_2\in[0,2\pi]$ for $N_s=26$ and  $\theta_1\in [0,2\pi],\theta_2\in[0,4\pi]$ for $N_s=30$. The average Chern number is $C_{mean}=-0.5$ for each state within a single period ($\theta_1\in [0,2\pi],\theta_2\in[0,2\pi]$). }
\label{fig:Chern_26_30}
\end{localgraphicspath}
\end{figure}

In Fig.\ref{fig:Chern_26_30}, we show energy spectrum under flux insertion and the many-body berry curvature for $N_s=26,30$ with $\gamma=5.5$. For those clusters, the two ground state momentum sectors will evolve into each other by flux insertion over a single period in both directions $\theta_1\in[0,2\pi]$ and $\theta_2\in[0,2\pi]$. In order to show the evolution of the two sectors, we stick to a single momentum sector but evolve over two periods. We find the spectrum gap is finite under flux insertion and the averaged Chern number is $C_{mean}=-0.5$ for each state within a single period, which is consistent with half filled FCIs. 

\begin{figure}
\begin{localgraphicspath}{{appendix_fig/}}
\includegraphics[width=0.9\columnwidth]{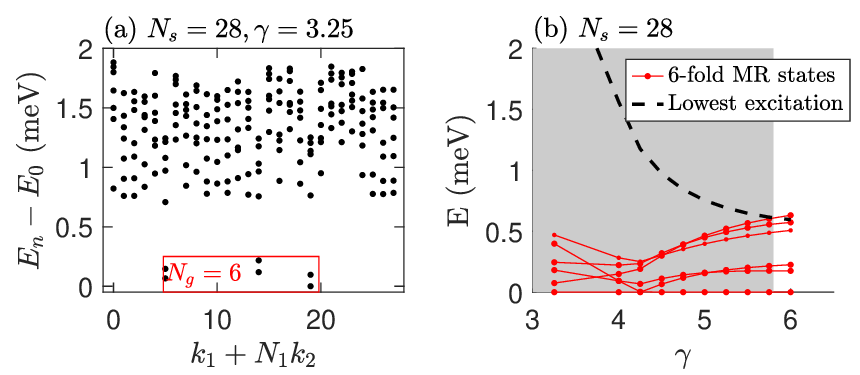}
\caption{Momentum resolved energy spectrum (a) and phase diagram for the half filled t1 band with $\epsilon=2$. }
\label{fig:spectrum_e1}
\end{localgraphicspath}
\end{figure}

\section{Results on half filled first top band}
In the main text, we showed results of Moore-Read states in first bottom band (b1). Here we provide further results to show that Moore-Read state can also be realized in first top band (t1). We take $\epsilon=2$ since the band width of t1 band is larger (Fig. 1 in main text). In Fig.\ref{fig:spectrum_e1} (a) one can also see $6$-fold quasi-degenerate states which lie in the same momentum sectors as in the case of b1 band.  In Fig.\ref{fig:spectrum_e1} (b) one can see finite gap of Moore-Read states up to $\gamma=5.8$.

\end{document}